\documentstyle[11pt]{article}

\newtheorem{thm}{Theorem}[section]

\begin{document}

\title{\bf  {Homothetic Motion in a Bianchi Type-I Model in Lyra Geometry}}
\author{{ Ragab M. Gad$^{1,2}$ \thanks{%
E-mail: ragab2gad@hotmail.com} }\\
\newline
{\it $^1$
 Mathematics Department, Faculty of Science, King
Abdulaziz University,}\\
 {\it  21589 Jeddah, KSA}
 \\
{\it $^2$ Mathematics Department, Faculty of Science, Minia University,}\\
 {\it   61915 El-Minia,  Egypt}
}

\date{\small{}}

\maketitle

\begin{abstract}
In this paper we study a homothetic vector field of a Bianchi type-I model based on
Lyra geometry. The cases when a displacement vector is function
of $t$ and when it is  constant are considered.  In both two cases we investigate the equation of state.
A comparison between the obtained results, using Lyra geometry, and that have
obtained previously in the context of General Relativity,
based on Riemannian geometry, will be given.
\end{abstract}
{\bf{Keywords}}: Bianchi type-I; homothetic vector field; Lyra geometry; matter collineations; baratropic equation of state.  \\

\setcounter{equation}{0}
\section{Introduction}

The theme of symmetries of space-time is nearly as old as the outset of the theory of General Relativity.
Symmetries play an important role in the study of space-time, because of their interest from both a geometric and a physical viewpoint.  Symmetries have been studied in the theory of General Relativity based on Riemannian geometry and in the theory of teleparallel gravity based on the Weitzenbock geometry \cite{12}.
In the theory of general relativity different kind of symmetries like isometry, homothetic, conformal, Ricci  collineations and matter collineations have  been extensively studied \cite{H04}-\cite{G03}. Also some of them have been studied in the theory of teleparallel gravity, over the past few years \cite{SM09}-\cite{S11}.\\
In the context of Lyra geometry \cite{L51,S52}, Gad and Alofi \cite{GA14} discussed the homothetic symmetry on plane symmetric Bianchi type-I cosmological model. They assumed that the displacement vector field is either function of the time coordinates $t$ or it is constant and found the homothetic vector field in both two cases. They proved that this model  is a generalization of the FRW model. They investigated the baratropic equation of state and showed that it is not satisfied as either the displacement vector is function of $t$ or it is constant.

In this paper, we will study the symmetries of Bianchi cosmological models based on  Lyra geometry, in particular the homothetic symmetry of a Bianchi type-I model relativistic cosmology.
The reason for this is that this symmetry playing a dominant role in the dynamics of Bianchi cosmological models \cite{RJ85,C03}.\\
The paper is organized as follows: In the next Section, we summarize some of the basic concepts of Lyra geometry, which will be used through this work. Section 3 deals with the model and evaluating the homothetic vector field. In Section 4, we discuss the matter collineations for the model under consideration and investigate the baratropic equation of state. Finally, in Section 5, concluding remarks are given.

\setcounter{equation}{0}
\section{The Lie derivative and homothetic equation}
 An $n$-dimensional Lyra manifold $M$ is a generalization to the Riemannian manifold \cite{L51}. For any point $p\in M$ we can define the coordinate system $\{x^\mu\}_{\mu=1}^{n}$. In addition to these coordinates there exist a gauge function $x^o =x^o(x^\mu)_{\mu=1}^n$, which together with $\{x^\mu\}_{\mu=1}^{n}$ forms the so called a {\it{reference system transformation}} $(x^o,x^\mu)_{\mu=1}^n$. In Lyra geometry the metric or the measure of length of displacement vector $\zeta^\mu =x^o dx^\mu$ between two points $p(x^\mu)_{\mu=1}^n$ and $q(x^\mu+dx^\mu)_{\mu=1}^n$ is given by absolute invariant under both gauge function and coordinate system
 $$
 ds^2=g_{\mu\nu}x^odx^\mu x^o dx^\nu,
 $$
 where $g_{\mu\nu}$ is a metric tensor as in Riemannian geometry.\\
 In Lyra geometry, a generalized affine connection  characterized not only by the Riemannian connection, $\big\{\,^\alpha_{\mu\nu}\big\}$, but also by a function $\phi_\mu$, which arises through gauge transformation, and it is given by
\begin{equation}\label{conn}
\Gamma^{\alpha}_{\,\,\mu\nu}=(x^o)^{-1}\big\{\,^\alpha_{\mu\nu}\big\} + \frac{1}{2}\big(\delta^\alpha_\mu\phi_\nu + \delta^\alpha_\nu\phi_\mu -g_{\mu\nu}\phi^\alpha\big),
\end{equation}
where $\phi$ is called a displacement vector field and satisfies $\phi^\alpha =g^{\alpha\beta}\phi_\beta$. We consider $\phi$ to be a timelike vector, where
\begin{equation}\label{phi}
\phi_\mu=(\beta(t),0,0,0).
\end{equation}
The covariant derivative for a vector field ${\bf{\eta}}$ can be defined as follows
\begin{equation}
\begin{array}{ccl}
\nabla_\mu\eta^\nu &=&\frac{1}{x^o}\partial_\mu\eta^\nu + \Gamma^{\nu}_{\,\,\mu\alpha}\eta^\alpha,\\
\nabla_\mu\eta_\nu &=&\frac{1}{x^o}\partial_\mu\eta_\nu - \Gamma^{\alpha}_{\,\,\mu\nu}\eta_\alpha.
\end{array}
\end{equation}
where $\Gamma^{\sigma}_{\mu\nu}$ is a Lyra connection given by equation (\ref{conn}).\\
For a covariant tensor of second rank the covariant derivative is
\begin{equation}\label{CD}
\nabla_\rho Q_{\mu\nu}=\frac{1}{x^o}Q_{\mu\nu,\rho}-\Gamma^{\sigma}_{\rho\nu}Q_{\mu\sigma}-\Gamma^{\sigma}_{\mu\rho}Q_{\sigma\nu}.
\end{equation}
The Lyra connection is metric preserving:
\begin{equation}\label{g}
\nabla_\nu g_{\alpha\beta}=\frac{1}{x^o}g_{\alpha\beta,\nu}-g_{\alpha\mu}\Gamma^\mu_{\,\,\beta\nu}-g_{\mu\beta}\Gamma^{\mu\,\,\alpha\nu}=0.
\end{equation}
In Riemannian geometry, the Lie derivative of any tensor with respect to a vector field can be expressed
through the covariant derivative of that tensor and the vector field. With this consideration, the definition of the Lie derivative of this tensor in Lyra geometry is
\begin{equation}\label{Lie0}
\pounds_\eta Q_{\mu\nu}=(\nabla_\rho Q_{\mu\nu})\eta^\rho + (\nabla_\mu\eta^\rho)Q_{\rho\nu} +(\nabla_\nu\eta^\rho)Q_{\mu\rho},
\end{equation}
where $\nabla$ is the covariant derivative as defined in (\ref{CD}).\\
Inserting equation (\ref{CD}) into the above equation, we get
$$
\pounds_\eta Q_{\mu\nu} =\eta^\rho(\frac{1}{x^o}Q_{\mu\nu,\rho} -\Gamma^\sigma_{\,\,\rho\nu}Q_{\mu\sigma}-\Gamma^\sigma_{\mu\rho}Q_{\sigma\nu})
+ Q_{\rho\nu}(\frac{1}{x^o}\eta^\rho_{,\mu} + \Gamma^{\rho}_{\,\,\mu\sigma}\eta^\sigma)
$$
\begin{equation}\label{Lie}
+ Q_{\mu\rho}(\frac{1}{x^o}\eta^\rho_{,\nu} +\Gamma^\rho_{\sigma\nu}\eta^\sigma).
\end{equation}
From  equation (\ref{Lie0}), using (\ref{g}), if $Q_{\mu\nu}=g_{\mu\nu}$, we have
$$
\pounds_\eta g_{\mu\nu}=g_{\rho\nu}\nabla_\mu\eta^\rho +g_{\mu\rho}\nabla_\nu\eta^\rho.
$$
As in Riemannian geometry if
\begin{equation}\label{HE}
\pounds_\zeta g_{\mu\nu}=g_{\rho\nu}\nabla_\mu\zeta^\rho +g_{\mu\rho}\nabla_\nu\zeta^\rho=2\psi g_{\mu\nu},
\end{equation}
where $\psi$ is a constant, then $\zeta$ is called {\it{homothetic vector field}} and the above equation is called {\it{homothetic equation}}. It is worth mention here if $\psi =0$, then equation (\ref{HE}) is called {\it{Killing equation}} and $\zeta$ is called a {\it{Killing vector field}}.

\setcounter{equation}{0}
\section{\bf{The model and homothetic vector field}}
A Bianchi type-I cosmological model of the Universe attracted considerable attention in the relativistic cosmology literature. It may be due to the fact that the model is the straightforward generalization of the flat Friedman-Robertson-Walker (FRW) model. It is also one of the simplest models of the anisotropic Universe that describes a homogeneous and spatially flat Universe. Moreover, This space-time has a different scale factor in each direction, thereby introducing an isotropic to the system, unlike the FRW space-time which has the same scale factor for each of the three spacial directions.\\
We consider Bianchi type-I metric in the form:
\begin{equation}\label{BI}
ds^{2}=dt^{2}-A^{2}(t)dx^{2}-B^{2}(x,t)dy^{2}-C^{2}(x,t)dz^{2},
\end{equation}
with the convention $(x^{0}=t$, $x^{1}=x$, $x^{2}=y$, $x^{3}=z)$,
$A$ is a function of $t$ only and $B$ and $C$ are functions of $x$ and $t$.\\
For a vector field ${\bf{\zeta}}=\zeta^\mu(t,x,y,z)_{\mu=1}^4$, the homothetic  equations (\ref{HE}) for the model (\ref{BI}), using (\ref{conn}) and apart from the factor $\frac{1}{x^o}$, i.e., we choose the normal gauge $x^o=1$, are reduced to the
following system of equations:
\begin{equation}\label{1}
\zeta^1_{,1} + \big(\frac{\dot{A}}{A} +\frac{1}{2}\beta\big)\zeta^0=\psi,
\end{equation}
\begin{equation}\label{2}
A^2\zeta^1_{,2}+B^2\zeta^2_{,1}=0,
\end{equation}
\begin{equation}\label{3}
A^2\zeta^1_{,3}+C^2\zeta^3_{,1}=0,
\end{equation}
\begin{equation}\label{4}
\zeta^0_{,1}-A^2\zeta^1_{,0}=0,
\end{equation}
\begin{equation}\label{5}
\zeta^2_{,2} +\frac{B^\prime}{B}\zeta^1 +\big(\frac{\dot{B}}{B} +\frac{1}{2}\beta\big)\zeta^0=\psi,
\end{equation}
\begin{equation}\label{6}
B^2\zeta^2_{,3} +C^2 \zeta^3_{,2}=0,
\end{equation}
\begin{equation}\label{7}
\zeta^0_{,2}-B^2\zeta^2_{,0}=0,
\end{equation}
\begin{equation}\label{8}
\zeta^3_{,3} +\frac{C^\prime}{C}\zeta^1 +\big(\frac{\dot{C}}{C} +\frac{1}{2}\beta\big)\zeta^0=\psi,
\end{equation}
\begin{equation}\label{9}
\zeta^0_{,3}-C^2\zeta^3_{,0}=0,
\end{equation}
\begin{equation}\label{10}
\zeta^0_{,0} + \frac{1}{2}\beta\zeta^0=\psi.
\end{equation}
Solving equation (\ref{10}) and using the result back into equations (\ref{4}), (\ref{7}) and (\ref{9}), we get
\begin{equation} \label{1-}
\zeta^0=[\psi\int{e^{\frac{1}{2}\int{\beta dt}}dt} +c_0]e^{-\frac{1}{2}\int{\beta dt}},
\end{equation}
$$
\zeta^1=F_1(x,y,z),
$$
$$
\zeta^2=F_2(x,y,z),
$$
$$
\zeta^3=F_3(x,y,z),
$$
where $c_0$ is a constant of integration and $F_1(x,y,z)$, $F_2(x,y,z)$ and $F_3(x,y,z)$ are arbitrary functions  which are to be determined.\\
Differentiating (\ref{1}) with respect to $t$ and using (\ref{10}) and (\ref{1-}), we have
\begin{equation} \label{2-}
\frac{\dot{A}}{A}+\frac{1}{2}\beta=\frac{a}{[\psi\int{e^{\frac{1}{2}\int{\beta dt}}dt} +c_0]e^{-\frac{1}{2}\int{\beta dt}}},
\end{equation}
where $a$  is a constant of integration. Substituting this result back into (\ref{1}), using equations (\ref{2}) and (\ref{3}), gives
\begin{equation} \label{3-}
\zeta^1=(\psi -a)x +c_1,
\end{equation}
where $c_1$ is a constant of integration.\\
Using the above results, we find from equations (\ref{2}), (\ref{3}) and (\ref{6}) that $\zeta^2$ depends only on $y$ and $\zeta^3$ depends only on $z$. Consequently, we get from equations (\ref{5}) and (\ref{8}) that
\begin{equation} \label{4-}
B(x,t)=C(x,t),
\end{equation}
 \begin{equation} \label{5-}
\zeta^2=\psi y +c_2,
\end{equation}
\begin{equation} \label{6-}
\zeta^3=\psi z +c_3,
\end{equation}
where $c_2$ and $c_3$ are constants of integrations.\\
Without loss of generality, we assume that $c_0 =c_1 =c_2 =c_3= 0$, therefore, from equations (\ref{1-}), (\ref{3-}), (\ref{5-}) and (\ref{6-}) we obtain the following homothetic vector field
\begin{equation} \label{HVF}
{\bf{\zeta}}=([\psi\int{e^{\frac{1}{2}\int{\beta dt}}dt} ]e^{-\frac{1}{2}\int{\beta dt}})\, \partial_t + (\psi - a)x\,\partial_x +\psi y\partial_y +\psi z\partial_z.
\end{equation}
All the above considerations lead to the following theorem:
\begin{thm}
In Lyra geometry, if a displacement vector is function of $t$, that is, $\beta=\beta(t)$,  a Bianchi type-I space-time described by metric (\ref{BI}) admits the homothetic vector field (\ref{HVF}) if
$$
B(x,t) = C(x,t),
$$
$$
\frac{\dot{A}}{A}+\frac{1}{2}\beta=\frac{a}{[\psi\int{e^{\frac{1}{2}\int{\beta dt}}dt}]e^{-\frac{1}{2}\int{\beta dt}}},
$$
and
\begin{equation}\label{bb}
(\psi-a)xB^{\prime\prime} + (\psi-a)B^\prime + (\dot{B}^\prime +\frac{1}{2}\beta B^\prime)([\psi\int{e^{\frac{1}{2}\int{\beta dt}}dt}]e^{-\frac{1}{2}\int{\beta dt}})=0.
\end{equation}
\end{thm}
{\bf{proof}}:\\
From the above calculations, putting $c_0=0$ in (\ref{1-}), the first and second relations are satisfied. Inserting the components of the homothetic vector field into (\ref{8}), using the second relation, and differentiating the result with respect to $x$, we obtain the  relation (\ref{bb}).\\
Now we discuss the case when the displacement vector is constant, that is, $\beta=$ constant.\\
From equation (\ref{1-}), we have
$$
\zeta^0 = (\frac{2\psi}{\beta}e^{\frac{1}{2}\beta t} +c_0)e^{-\frac{1}{2}\beta t}.
$$
Hence in this case, putting $c_0=0$ in (\ref{1-}), the homothetic vector field is
\begin{equation} \label{HVF-}
{\bf{\zeta}}= \frac{2\psi}{\beta}\, \partial_t + (\psi - a)x\,\partial_x +\psi y\,\partial_y +\psi z\,\partial_z.
\end{equation}
Also, from equation (\ref{2-}), we have
$$
\frac{\dot{A}}{A}+\frac{1}{2}\beta=\frac{a}{(\frac{2\psi}{\beta}e^{\frac{1}{2}\beta t} )e^{-\frac{1}{2}\beta t}},
$$
Integrating this equation, we get
\begin{equation} \label{2-b}
A(t)= a_0\big( \frac{2\psi}{\beta}e^{\frac{1}{2}\beta t}\big)^{\frac{a}{\psi}}e^{-\frac{1}{2}\beta t},
\end{equation}
where $a_0$ is a constant of integration.\\
Due to the previous considerations we can state the following:
\begin{thm}
In Lyra geometry, if a displacement vector is constant,  a Bianchi type-I space-time described by metric (\ref{BI}) admits the homothetic vector field (\ref{HVF-}) if
$$
B(x,t) = C(x,t),
$$
$$
A(t)= a_0\big( \frac{2\psi}{\beta}e^{\frac{1}{2}\beta t}\big)^{\frac{a}{\psi}}e^{-\frac{1}{2}\beta t},
$$
and
$$
(\psi-a)xB^{\prime\prime} + (\psi-a)B^\prime + (\dot{B}^\prime +\frac{1}{2}\beta B^\prime)((\frac{2\psi}{\beta}e^{\frac{1}{2}\beta t})e^{-\frac{1}{2}\beta t})=0..
$$
\end{thm}
It is interesting to note that, if one consider the particular class of models (\ref{BI}) for which the scale factors $B(t,x)$ and $C(t,x)$ are functions of $t$ only, this allow us to compare our results with that obtained in the theory of General Relativity. In this case and after a straightforward calculations, using the above procedure for finding the components of the homothetic vector field, we get
\begin{enumerate}
\item The components $\zeta^0$ and $\zeta^1$ of the homothetic vector field are the same as given, respectively, in equations (\ref{1-}) and (\ref{3-}).
\item From equations (\ref{5}) and (\ref{8}), we get, respectively
\begin{equation}\label{4--}
\zeta^2 =(\psi -b)y + b_1,
\end{equation}
\begin{equation}\label{5--}
\zeta^3 =(\psi -d)z + d_1,
\end{equation}
where $b_1$ and $d_1$ are constants of integration and the constants $b$ and $d$ are given from the following relations, after using equations (\ref{1-}) and (\ref{3-}) in equations (\ref{5}) and (\ref{8})
\begin{equation}\label{2--}
\frac{\dot{B}}{B}+\frac{1}{2}\beta=\frac{b}{[\psi\int{e^{\frac{1}{2}\int{\beta dt}}dt} +c_0]e^{-\frac{1}{2}\int{\beta dt}}},
\end{equation}
\begin{equation}\label{2---}
\frac{\dot{C}}{C}+\frac{1}{2}\beta=\frac{d}{[\psi\int{e^{\frac{1}{2}\int{\beta dt}}dt} +c_0]e^{-\frac{1}{2}\int{\beta dt}}}.
\end{equation}
We note that in this case $B(t) \neq C(t)$.
\end{enumerate}
From equations (\ref{1-}), (\ref{3-}), (\ref{4--}) and (\ref{5--}), putting $c_0=c_1=b_1=d_1=0$, we obtain the following homothetic vector field
\begin{equation} \label{HVF1}
{\bf{\zeta}}= [\psi\int{e^{\frac{1}{2}\int{\beta dt}}dt}]e^{-\frac{1}{2}\int{\beta dt}}\, \partial_t + (\psi - a)x\,\partial_x +(\psi-b) y\,\partial_y +(\psi-d) z\,\partial_z.
\end{equation}
Putting $a=b=d$ in equations (\ref{2-}) , (\ref{2--}) and (\ref{2---}), by integrating them and equally between the integration constants, we get
\begin{equation}\label{abd}
A(t)=B(t)=C(t)=a_0\exp({\int{(\frac{a}{[\psi\int{e^{\frac{1}{2}\int{\beta dt}}dt}]e^{-\frac{1}{2}\int{\beta dt}}}-\frac{1}{2}\beta)dt}}).
\end{equation}
Now the following result has been
established:
\begin{thm}
In Lyra geometry, if a displacement vector is function of $t$, a Bianchi type-I space-time  with scalar factors are functions of $t$ only, is a FRW cosmological model with scalar factor behaves as
$$
a^2_0\exp{({\int{(\frac{2a}{[\psi\int{e^{\frac{1}{2}\int{\beta dt}}dt}]e^{-\frac{1}{2}\int{\beta dt}}}-\beta)dt}})}.
$$
\end{thm}
In the case $\beta$= constant, the above homothetic vector field becomes
\begin{equation} \label{HVF2}
{\bf{\zeta}}=\frac{2\psi}{\beta}\, \partial_t + (\psi - a)x\partial_x +(\psi-b) y\partial_y +(\psi-d) z\partial_z.
\end{equation}
Integrating the relations (\ref{2--}) and (\ref{2---}), we get, respectively
\begin{equation}\label{2-2}
B(t)= b_0\big( \frac{2\psi}{\beta}e^{\frac{1}{2}\beta t}+c_0\big)^{\frac{a}{\psi}}e^{-\frac{1}{2}\beta t},,
\end{equation}
\begin{equation}\label{2-3}
C(t)=d_0\big( \frac{2\psi}{\beta}e^{\frac{1}{2}\beta t}+c_0\big)^{\frac{a}{\psi}}e^{-\frac{1}{2}\beta t},,
\end{equation}
where the non-zero constants $b_0$ and $d_0$ are constants of integration.\\
As in the above theorem the following result has been
established:
\begin{thm}
In Lyra geometry, if a displacement vector is constant, a Bianchi type-I space-time  with scalar factors are functions of $t$ only, is a FRW cosmological model with scalar factor behaves as follows
$$
A^2(t)= B^2(t)=C^2(t)=a^2_0\big( \frac{2\psi}{\beta}e^{\frac{1}{2}\beta t}+c_0\big)^{\frac{2a}{\psi}}e^{-\beta t}.
$$
\end{thm}
To compare our results with that obtained in the theory of General Relativity based on Riemannian geometry, we put $\beta =0$ and assume $\psi =1$ for simplicity. From theorem (3.1), $\beta=\beta(t)$, the homothetic vector becomes
$$
{\bf{\zeta}}=t\, \partial_t + (1 - a)x\,\partial_x + y\,\partial_y + z\,\partial_z,
$$
but from theorem (3.2), $\beta=$const., we can not compare the results with that obtained in the theory of General Relativity, because in this case the component $\zeta^0$ tends to infinity when $\beta=0$. \\
From theorem (3.3), $\beta =\beta(t)$ and the scalars factors are functions of $t$ only,  the homothetic vector field takes the following form
\begin{equation} \label{HVF3}
{\bf{\zeta}}=t\, \partial_t + (1 - a)x\,\partial_x +(1-b) y\,\partial_y +(1-d) z\,\partial_z,
\end{equation}
and from equations (\ref{2-}), (\ref{2-2}) and (\ref{2-3}), the scale factors behave as follows
\begin{equation}\label{2-a}
A(t)= a_0t^a,
\end{equation}
\begin{equation}\label{2-b}
B(t)=b_0t^b,
\end{equation}
\begin{equation}\label{2-c}
C(t)=d_0t^d.
\end{equation}
Now we conclude that the homothetic vector field given by (\ref{HVF3}) and with the above constrains on the scalar factors  is  the same as obtained in the theory of General Relativity \cite{Be12,Be13}.\\
From theorem (3.4), $\beta=$ const., when we assume that $\beta=0$ due to both $\zeta^0$ and the scalar factors tend to infinity. Therefore, in this case we can not compare the results obtained using Lyra geometry with that obtained in General Relativity using Riemannian geometry.

\setcounter{equation}{0}
\section{Matter collineation}
In recent years, much interest has been shown in the study of matter collineations (MCs) (see for instance \cite{S01a}, \cite{S01b}, \cite{S05}, \cite{TA04})\\
A vector field along which the Lie derivative of energy-momentum tensor vanishes is called an MC, that is
$$
\pounds_XT_{\mu\nu}=0,
$$
where ${\bf{X}}$ is the symmetry or collineation vector. Also, assuming the Einstein's field equations, a vector ${\bf{X}}$ generates an MC if $\pounds_XG_{\mu\nu}=0$.\\
It is obvious that the symmetries of the metric tensor (isometries) are also symmetries of the Einstein tensor $G_{\mu\nu}$, but this is not necessarily the case for the symmetries of the Ricci tensor (Ricci collineations) which are not, in general, symmetries of the Einstein tensor.\\
If ${\bf{X}}$ is a Killing vector (or a homothetic vector) then $\pounds_XT_{\mu\nu}=0$, thus every isometry is also an MC but converse is not true, in general.\\
In the theory of General Relativity Cahill and Taub \cite{CT71} and Bicknell and Henriksen [15] (see also \cite{OP90}) pointed out, if
the matter field is a perfect fluid, then the only baratropic equation of state which is
compatible with self-similarity (characterized by the existence of homothetic vector field) is of the form
\begin{equation}\label{bara}
\rho=kp,
\end{equation}
where $\rho$ is the total energy density, $p$ is the pressure and $k$ is a constant in the range
$0\leq k \leq 1$. This equation of state is nevertheless physically consistent in the whole range
of $k$. When $k = 0$, the above equation describes dust and $k =1/3$ gives the equation of state for
radiation.\\
It is of interesting to assume that the matter of field in space-time under consideration is represented by a perfect fluid,
 that is, the energy-momentum tensor is defined by
\begin{equation}\label{EMT}
T_{\mu\nu}=(\rho +p)u_{\mu}u_\nu-pg_{\mu\nu},
\end{equation}
 where, for the space-time (\ref{BI}), $u^\mu=(1, 0, 0, 0)$, $u^\mu u_\mu =1$, is the four-velocity vector field.
This allows us to compare our results with that the results obtained previously in the theory of General Relativity.\\
Using the components of the homothetic vector (\ref{HVF}) in the definition of Lie derivative (\ref{Lie}) for the energy-momentum tensor, $T_{\mu\nu}$, (\ref{EMT}), then for the space-time (\ref{BI}) the equation $\pounds_\zeta T_{\mu\nu}=0$, reduces to the following equations
\begin{equation}\label{MCV}
\begin{array}{ccl}
\zeta^0 T_{00,0} +\zeta^1T_{00,1} +2T_{00}\zeta^0_{,0} &=& 0,\\
\zeta^0 T_{11,0} +\zeta^1T_{11,1} +2T_{11}\zeta^1_{,1} &=& 0,\\
\zeta^0 T_{22,0} +\zeta^1T_{22,1} +2T_{22}\zeta^2_{,2} &=& 0,\\
\zeta^0 T_{33,0} +\zeta^1T_{33,1} +2T_{33}\zeta^3_{,3} &=& 0.
\end{array}
\end{equation}
Now, we assume that both the pressure and density for the space-time under consideration are functions of $t$ only. Therefore, for the space-time (\ref{BI}), the first equation in (\ref{MCV}), using equations (\ref{EMT}) and (\ref{1-}) and assuming $c_0=0$, reduces to
\begin{equation}\label{density}
\rho = \frac{r_1}{\big( [\psi\int{e^{\frac{1}{2}\int{\beta dt}}dt}]e^{-\frac{1}{2}\int{\beta dt}} \big)^2},
\end{equation}
where $r_1$ is a constant of integration.\\
From the second equation of the system (\ref{MCV}), taking into account equation (\ref{1-}), we have
\begin{equation}\label{pressure}
p=r_2 \exp{\big(\int{(-\frac{(2\psi -a)}{[\psi\int{e^{\frac{1}{2}\int{\beta dt}}dt}]e^{-\frac{1}{2}\int{\beta dt}}}+\frac{1}{2}\beta)}dt\big)},
\end{equation}
where $r_2$ is a constant of integration.\\
From the third equation of the system (\ref{MCV}), using equation (\ref{5}), we get
\begin{equation}\label{pressure-}
p=r_3 \exp{\big(\int{(\frac{-2\psi}{[\psi\int{e^{\frac{1}{2}\int{\beta dt}}dt}]e^{-\frac{1}{2}\int{\beta dt}}}+\frac{1}{2}\beta)}dt\big)},
\end{equation}
where $r_3$ is a constant of integration.\\
Assuming $r_2=r_3$, equations (\ref{pressure}) and (\ref{pressure-}) imply that $a=0$. Consequently, a homothetic vector which is an MC vector, using theorem (3.1) has the following form
\begin{equation} \label{HVF-MC}
{\bf{\zeta}}=([\psi\int{e^{\frac{1}{2}\int{\beta dt}}dt} ]e^{-\frac{1}{2}\int{\beta dt}})\, \partial_t + \psi x\,\partial_x +\psi y\partial_y +\psi z\partial_z.
\end{equation}
 In this case
 $$
 A(t)=a_0e^{-\frac{1}{2}\int{\beta dt}}.
 $$
From equations (\ref{density}) and (\ref{pressure-}) it is not easy to find relation between $\rho$ and $p$, therefore, we have the following interesting result:\\
In Lyra geometry, if a displacement vector is function of $t$, then the baratropic equation of state (\ref{bara}) is not satisfied, in contrast of General Relativity.\\
Now, we discuss the case when the displacement vector is constant, that is, $\beta=$ constant, in this case the density and the pressure, respectively, are
\begin{equation}\label{density1}
\rho = \frac{r_1\beta^2}{4\psi^2},
\end{equation}
\begin{equation}\label{pressure1}
p=r_2 e^{-\frac{1}{2}\beta t},
\end{equation}
It is not easy to find relation between $\rho$ and $p$ from (\ref{density1}) and (\ref{pressure1}). Therefore, the baratropic equation of state can not be satisfied in the case when $\beta=$ const.\\
Now, we conclude that when we study the space-time using Lyra geometry, so the equation of state is not satisfied, whether the displacement vector is function in time or it is constant.\\

Now, we will compare our results with that obtained in the theory of General Relativity, to do this, we put $\beta =0$ and assume $\psi =1$, for simplicity. In the case when $\beta=\beta(t)$, from equations (\ref{density}) and (\ref{pressure-}),  the  density and pressure take, respectively, the following forms
\begin{equation}\label{density2}
\rho = \frac{r_1}{t^2},
\end{equation}
\begin{equation}\label{pressure2}
p=\frac{r_2}{t^{2}}.
\end{equation}
To satisfy the equation of state (\ref{bara}), we  put $\frac{r_1}{r_2}=k$ in the above equations, that is, for the space-time under consideration the baratropic  equation of state is satisfied  in the theory of General Relativity.\\
We note that in the case when $\beta=$ const., from equation (\ref{density1}), the density $\rho$ tends to zero. This is contrary to the assumption that the space-time (\ref{BI}) is represented by a perfect fluid. Therefore, in this case we can not compare our results with that obtained in the theory of General Relativity.

\section{Discussion and conclusion}
This work is devoted to studying the symmetries, in particular homothetic symmetry, of a Bianchi type-I space-time in the framework of Lyra geometry. We have focussed in this kind of symmetries since a space-time admitting it is stable from the dynamical system point and therefore is in important from the physical one. For a Bianchi type-I space-time, we have obtained the homothetic vector field in the context of Lyra geometry, by assuming that the displacement vector field is either function of the coordinate $t$ or it is constant. In these two cases, we have proved that a Bianchi type-I model is a generalization of the FRW model, this result agrees with that  obtained in the theory of General Relativity. In this paper, we have classified the space-time (\ref{BI}) according to admitting a homothetic vector field. Moreover, we have explained the matter collineation symmetry, by assuming that the matter field is represented by perfect fluid. This allowed us to obtained the homothetic vector field which is a matter collineation vector field, and gave  the possibility to investigate the baratropic equation of state (\ref{bara}). In the context of Lyra geometry, we found that this equation is not satisfied either a displacement vector is function of $t$ or it is constant. The only classifications of the space-time under consideration which lead to satisfy the baratropic equation of state in the General Relativity, by assuming $\beta =0$, are the solutions for which a displacement vector field is function of $t$. This work can be extended to other space-times to make a comparison of symmetries of cosmological models in the framework of Lyra geometry and in the theory of General Relativity.
\section*{Acknowledgments}
This work was funded by the Deanship of Scientific Research (DSR), King Abdulaziz University, Jeddah, under grant No. (965-017-D1434).
The author, therefore, acknowledge with thanks DSR technical and financial support.

\end{document}